\documentstyle[prd,aps,epsf]{revtex}

\newcommand{\Abar}{\not{\!{\!A}}}
\newcommand{\Bbar}{\not{\!{\!B}}}
\newcommand{\Pbar}{\not{\!{\!P}}}
\newcommand{\Sbar}{\not{\!{\!S}}}
\newcommand{\pabar}{\not{\!\partial}}
\newcommand{\Od}{{\cal O}}

\newcommand{\Tr}{\mbox{Tr}}
\newcommand{\im}{\mbox{Im}}

\newcommand{\Dbar}{\not{\!{\!D}}}
\newcommand{\qbar}{\not{\!q}}
\newcommand{\pbar}{\not{\!p}}

\newcommand{\re}{\mbox{Re}}

\begin{document}
\draft

\input epsf \renewcommand{\topfraction}{0.8}
%@twocolumnfalse\endcsname

\title{Particle production from axial fields}
\author{Antonio L. Maroto \thanks{E-mail:alm@star.cpes.susx.ac.uk } }
\address{Astronomy Centre, 
               University of Sussex,
               Falmer, 
               Brighton,
               U.K.
               BN1 9QJ}
\date{\today}
\maketitle
\begin{abstract}
We study the production of massive fermions in arbitrary vector and
axial-vector classical backgrounds using  effective action techniques.
A perturbative calculation shows the different features of each field and in
particular it is seen that pure temporal axial fields can produce particles
whereas it is not possible for a pure vector background. We also analyze
from a non-perturbative point of view a particular configuration 
with constant electric and axial fields and
show that the presence of the axial background inhibits the production from
the electric field. 
\end{abstract}

%\vskip2pc]

%%%%%%%%%%%%%%%%%%%%%%%%%%%%%%%%%%%%%%%%%%%%%%%%%%%%%%%%%%%%%%%%%%%%%%%

\section{ Introduction }
The production of particles from classical backgrounds has become
a very active area of research in the last years. We can find  it  in 
 numerous and disconnected fields of physics such as
cosmology \cite{BiDa,Parker}, heavy-ion collision or
even plasma physics \cite{ion}.  The pioneer work of 
Schwinger \cite{Schwinger}  mainly 
focused on the
production of electron-positron pairs by strong electrostatic fields.
Since then  many
other sources of particles creation have been studied in the literature.
Thus for example we can mention, time dependent gravitational fields 
\cite{Parker}, varying Planck
mass models \cite{gio}, compactification of extra dimensions 
\cite{Verdaguer}, dilaton field
\cite{gio}, inflaton field,  etc.  
The original Schwinger's work was based on the proper-time technique for 
the evaluation of the
effective action. This method allowed him  
to obtain an exact non-perturbative result
for the total number of particles produced.
However, although some other electromagnetic configurations have been studied
\cite{Brezin} 
and  some particular cases have also been  solved \cite{Naroz}, 
in general
the number of models for which exact results can be obtained is
very limited.  

In spite of its generalized use, it is probably in cosmology where particle 
production
has been applied more extensively. 
Thus for instance,  it is believed that the presence of any
small anisotropy in the early universe could have been erased very effectively
by particle emssion processes \cite{zeldovich}. The 
theory of reheating after inflation is also based on the resonant production
of particles due to the oscillations of the inflaton field \cite{Linde}.
The reheating of fermions has  been considered in \cite{Green}.  
In addition,
the formation of large scale structures in the early universe is closely
related to the generation from vacuum fluctuation of small seed density
inhomogeneities that due to the gravitational instability grew to give
rise to the current galactic structure. The generation of the vacuum  
fluctuations can be studied in much the same way as the production
of scalar particles in a Robertson-Walker background \cite{bran}. 
Recently the problem of particle production has also been extended to
the area of the string cosmology in which, apart from the gravitational  
background, there is also an additional scalar field, the dilaton, that
can also give rise to the generation of particles \cite{gio,string}.

In general, the problem of calculating  particle production can be 
approached in two different ways. On one hand, the Bogolyubov technique
\cite{BiDa} 
that allows the calculation of the spectrum of the particle produced. 
It is based on the resolution of the harmonic-oscillator equation
with variable frequency and only in some particular cases permits 
the derivation
of exact results. On the other hand we have the already mentioned
 effective action
technique that allows to obtain the total number of particles produced 
in a much simpler way \cite{Schwinger,hartle,DoMa2}, although the difficulties
in finding exact results are also present.

Most of the existing works about the creation of particles from
vacuum fluctuations  concentrate in the
production of boson fields in the presence of  scalar, vector or gravitational
backgrounds. In this paper we will study a different source for the production 
of fermions, it is the presence of
general vector and axial-vector backgrounds. We will thus extend the 
Schwinger's work by including the effects of a non-vanishing axial field. 
Classical axial backgrounds appear naturally in modern theories of
gravity such as supergravity \cite{sugra} or in low-energy string effective 
actions \cite{strings,Tseytlin}. Both theories contain torsion
(or axion) fields as a fundamental ingredient and, in fact, recently several
solutions  with non-vanishing axial
fields have been found in the context of string cosmology \cite{copeland,shap}.
On the other hand, the idea of modifying General Relativity by introducing
an arbitrary metric connection with torsion is an old one \cite{cartan}, and
in some sense quite natural from the point of view of the gauge theories
of gravity \cite{Kibble}. 
 This torsion field
is coupled minimally to fermions by means of its pseudotrace, thus providing
a new mechanism for the production of particles. In fact in this gravitational
theories, torsion would be the dominant mechanism for the production
of massless fermions in cosmological Robertson-Walker backgrounds. 
This is due to the fact
 that when gravity is minimally coupled to masless fermions, the theory is 
conformally invariant. This implies the well known result of absence 
of particle production. However, the presence of additional fields, such as 
torsion or metric anisotropies breaks that invariance.  
Our work will be based on the effective action method, 
first from  a perturbative
point of view and then we will study a particular case in which
a non-perturbative calculation is viable.

The paper is organized as follows: in section 2 we introduce the lagrangian
for the model and also give a brief introduction to the effective action
technique. In section 3, we perform the perturbative calculation and 
compare the result with the pure vector case. Section 4 is devoted to the
particular case with constant electric and axial field and a non-perturbative
result is obtained in the limit of small axial fields and in section
5 we give the main conclusions of the work. Finally we have included
an Appendix with some useful formalae of standard perturbation
theory in quantum mechanics.

\section{Model lagrangian and the effective action method}

We will consider the following interaction lagrangian for massive
fermions minimally coupled to abelian vector and axial-vector fields. 
For simplicity we will use the left-right notation and at the end
we will recover the vector-axial fields:
\begin{eqnarray}
{\cal L}=\overline \psi(i\Dbar-m+i\epsilon)\psi
\label{lag}
\end{eqnarray}
where $i\Dbar=i\gamma^\mu(\partial_\mu+iA_\mu P_L+iB_\mu P_R)$. As usual the
$+i\epsilon$ factor is introduced to ensure the convergence of the 
path integral and the left and right projectors are defined as:
$P_L=(1-\gamma_5)/2$ and $P_R=(1+\gamma_5)/2$. We will use
the chiral representation for the Dirac matrices in which $\gamma_5$
is diagonal. The coupling constants are
included in the own fields. 

Let us now introduce the effective action (EA) 
for the $A_\mu$ and $B_\mu$ fields
that is obtained after integrating the fermions out:
\begin{eqnarray}
\langle 0,t\rightarrow \infty\vert 0, t\rightarrow -\infty\rangle &=&
Z[A,B]=e^{iW[A,B]}=N\int d\psi d\overline\psi e^{i\int d^4x {\cal L}}\nonumber
\\ 
&=&N\det (i\Dbar -m +i\epsilon)
\label{ea}
\end{eqnarray}
Here, $\vert 0, t\rightarrow \pm \infty\rangle$ denote the initial and final
vacuum states that in general will be different due to the presence of
the external sources. $N$ is a normalization constant that is taken
as usual in such a way that $Z[0,0]=1$, this will allow to discard the vacuum
divergences as shown below.

The EA will be in general a complex non-local functional in the external 
fields. Its real part will contain the divergences that will be renormalized
by adding suitable local counterterms to the action. The imaginary part
will be finite and will contain the information about the particle 
production probabilities. In fact, the probability that the vacuum
remains stable is given by 
$\vert\langle 0,t\rightarrow \infty\vert 0, t\rightarrow 
-\infty\rangle\vert ^2$. Therefore the probability that the vacuum decays
by particle emission will be given by:
\begin{eqnarray}
P=1-\vert\langle 0,t\rightarrow \infty\vert 0, t\rightarrow 
-\infty\rangle\vert ^2=1-e^{-2\; Im W[A,B]}\simeq 2\;\im W[A,B]
\end{eqnarray}
When we only have a vector field such that its corresponding electric field
is constant, the probability density per unit time and unit volume 
$p$ can be obtained exactly and the result
is given by \cite{Schwinger}:
\begin{eqnarray}
p=\frac{e^2E^2}{4\pi^3}\sum_{n=1}^\infty \frac{1}{n^2}e^{-\frac{m^2n\pi}{eE}}
\label{eres}
\end{eqnarray}
However for non-constant electromagnetic fields or when the axial part
is switched on, the computation becomes very involved and it is neccessary to
rely on some perturbative method.

Before concluding this section we will mention that although the notion 
of torsion appears in strings and supergravity in slightly different 
ways, 
both can be interpreted as the antisymmetric part of the affine connection 
in pseudo-Riemannian geometry \cite{Tseytlin}. Thus, if the  components 
of the metric 
connection are:  
$\hat \Gamma^{\lambda}_{\;\;\mu\nu}$, its  
antisymmetric part:  
$T_{\;\;\mu\nu}^{\lambda}=\hat\Gamma_{\;\;\mu\nu}^{\lambda}
-\hat\Gamma_{\;\;\nu\mu}^{\lambda}$
is known as the torsion tensor. By means of the Einstein equivalence
principle, it is now possible to minimally couple  torsion to
fermion fields \cite{shapir,DoMa1}, one gets:
\begin{eqnarray}
{\cal L}=\overline\psi i\gamma^\mu\left(\partial_\mu+\Omega_\mu+\frac{i}{8}S_\mu
 \gamma_5\right)\psi 
\end{eqnarray}
where  $S_{\rho}=\epsilon_{\mu\nu\lambda\rho}T^{\mu\nu\lambda}$ is the
torsion pseudotrace and $\Omega_\mu$ the spin-connection. 

\section{Perturbative method}

In this section we present the evaluation of the EA in (\ref{ea})
as an expansion in the external fields, i.e an expansion in coupling
constants. Let us start by writting:
\begin{eqnarray}
W[A,B]=i\Tr\;\log\left(i\Dbar-m+i\epsilon\right)
\end{eqnarray}
That can formally be expanded as:
\begin{eqnarray}
W[A,B]=i\sum_{k=1}\frac{(-1)^k}{k}\Tr\left((i\pabar-m)^{-1}
(\Abar P_L+\Bbar P_R)\right)^k
\end{eqnarray}
where the Dirac propagator is defined as usual by:
\begin{eqnarray}
(i\pabar-m)^{-1}_{xy}=\int d\tilde qe^{-iq(x-y)}\frac{\qbar+m}{q^2-m^
2+i\epsilon}
\end{eqnarray}
The functional traces $\Tr$ are evaluated in  dimensional
regularization with $D=4-\epsilon$ and $d\tilde q=\mu^\epsilon d^Dq/(2\pi)^D$.
The lowest order contribution in the expansion is given by the two-point
terms, i.e:
\begin{eqnarray}
W[A,B]^{(2)}=\frac{i}{2}\int d^4x d^4y d\tilde p d\tilde q\frac{\qbar+m}{q^2-m^2}
e^{-iq(x-y)}(\Abar_yP_L+\Bbar_yP_R)\frac{\pbar+m}{p^2-m^2}
e^{-ip(y-x)}(\Abar_xP_L+\Bbar_xP_R)
\end{eqnarray}
Expanding these terms and defining $k=q-p$, we obtain the following expression:
\begin{eqnarray}
W[A,B]^{(2)}&=&2i\int d^4x d^4y d\tilde k d\tilde q
\frac{e^{-ik(x-y)}}{(q^2-m^2)((q-k)^2-m^2)}\left(q^\mu A_\mu^y\frac{(q-k)^\rho}
{2}A_\rho^x \right.\nonumber \\
&-&\left.q^\mu\frac{(q-k)_\mu}{2}A^\nu_y A^x_\nu+q^\mu A_\mu^x A_\nu^y
\frac{(q-k)^\nu}{2}+\frac{m^2}{2}g^{\mu\nu}A_\mu^y B_\nu^x + (A\rightarrow B)
\right)
\end{eqnarray}
All the integrals, except for that involved in the term proportional to
$m^2$, can now be reduced to a common form that is evaluated in
dimensional regularization:
\begin{eqnarray}
\int d\tilde q \frac{q^\alpha (q-k)^\beta}{(q^2-m^2)((q-k)^2-m^2)}=
\frac{i}{(4\pi)^{D/2}}\left(k^\alpha k^\beta \Gamma (2-D/2)\int_0^1
dt \frac{t^2-t}{{\cal D}^{2-D/2}}-\frac{g^{\alpha\beta}}{2}\Gamma(1-D/2)
\int_0^1 dt {\cal D}^{D/2-1}\right)
\end{eqnarray}
with ${\cal D}=m^2-k^2 t(1-t)$. The integral proportional to  $m^2$ 
is nothing but:
\begin{eqnarray}
\int d\tilde q \frac{1}{(q^2-m^2)((q-k)^2-m^2)}=\frac{1}{(4\pi)^{D/2}}
\Gamma(2-D/2)\int_0^1 dt {\cal D}^{D/2-2}
\end{eqnarray}
The imaginary part of $W[A,B]$ can be easily
extracted from the integrals in the Feynman parameter $t$. They give rise to:
\begin{eqnarray}
\im \int_0^1 dt (t^2-t)\log(m^2-k^2t(1-t)-i\epsilon)&=&
\int_0^1 dt (t^2-t)\left\{\begin{array}{c}
0,\; m^2>k^2t(1-t)\\
-\pi/2, \;  m^2=k^2t(1-t)\\
-\pi,\; m^2<k^2t(1-t)
\end{array}\right.\nonumber \\
=\frac{\pi}{4}\sqrt{1-\frac{4m^2}{k^2}}\left(\frac{2}{3}+\frac{4m^2}
{3k^2}\right),\;\;\frac{k^2}{m^2}>4
\end{eqnarray}  
and in a similar fashion we obtain:
\begin{eqnarray}
\im \int_0^1 dt(m^2-k^2t(1-t)) \log(m^2-k^2t(1-t)-i\epsilon)&=&
-\pi\sqrt{1-\frac{4m^2}{k^2}}\left(\frac{2}{3}m^2-\frac{k^2}{6}\right)
\end{eqnarray}
and
\begin{eqnarray}
\im \int_0^1 dt \log(m^2-k^2t(1-t)-i\epsilon)
=
-\pi\sqrt{1-\frac{4m^2}{k^2}}\left(\frac{2}{3}m^2-\frac{k^2}
{6}\right),\;\;\frac{k^2}{m^2}>4 
\end{eqnarray}
Putting all the contributions together and changing to the vector and
axial-vector fields defined by $V=B+A$ and $S=B-A$ respectively, we obtain
the final result for the imaginary part in terms of the Fourier transformed
fields:
\begin{eqnarray}
\im W^{(2)}[A,B]&=&\frac{1}{8\pi^2}\int d\tilde k \theta(k^2-4m^2)
\sqrt{1-\frac{4m^2}{k^2}}\left(-\frac{1}{6}\left(1+\frac{2m^2}{k^2}\right)
(F_{\mu\nu}(k)F^{\mu\nu}(-k)\right.\nonumber \\
&+&\left.S_{\mu\nu}(k)S^{\mu\nu}(-k))+2m^2 S_\mu(k)
S^\mu(-k)\right)
\label{pert}
\end{eqnarray}
We see that unlike the vector case, the axial contribution to the imaginary 
part has an additional term proportional to $m^2 S^2$. This term is 
prohibited by gauge invariance in the vector case, however as it is well-known
it may appear in axial theories with massive fermions since those theories
violate the corresponding gauge invariance. In fact studying the divergences
that appear in the model we see that they are proportional to the following
operators: $F_{\mu\nu}F^{\mu\nu}$,  $S_{\mu\nu}S^{\mu\nu}$ and 
$m^2 S_\mu S^\mu$ \cite{shapiro,DoMa1}. 
The $S^4$ operator although having the same dimension does not
contribute to the divergent part.  
Therefore it is only neccesary to introduce a kinetic and a mass
counterterms for the axial field in order to render the theory finite.

Since the integrand of  imaginary part has to be understood as a probability
density, it is important to verify that it is always  positive. As far as
 $k^2>4m^2$, due to the step function present
in (\ref{pert}) it is possible to find a reference frame in which $\vec k=0$. 
Then we have:
\begin{eqnarray}
&\;&\frac{1}{3}\left(1+\frac{2m^2}{k_0^2}\right)\left(k_0 S^0(k) 
k_0 S^0(-k)-k_0^2
S_\mu(k)S^\mu(-k)+S\rightarrow V\right)+2m^2 S_\mu(k)S^\mu(-k)\nonumber \\
&=&\frac{1}{3}\left(1+2\frac{m^2}{k_0^2}\right)
(\vert S_i \vert^2+\vert V_i \vert^2)
+2m^2(\vert S_0 \vert^2-\vert S_i \vert^2)=
\frac{1}{3}(k_0^2-4m^2)\vert S_i \vert^2+\frac{1}{3}(k_0^2+2m^2)
\vert V_i\vert^2
+2m^2\vert S_0 \vert ^2 \geq 0
\end{eqnarray}
From this expression we can extract some of the different features
of the production from vector and axial fields. First we see that
if the axial field is purely spatial in the above reference frame, i.e., 
$S_0=0$ and we choose $S_i(k)=V_i(k)$, then the production from pure 
axial fields is always
supressed with respect to the pure vector case. However when $S_0(k)=V_0(k)$
and $S_i=V_i=0$ then, whereas there is no production in the vector
case, it is possible to create particles in the axial one. In the massless
case both fields give rise to the same amount of particles.

\section{Constant electric and axial fields: a non-perturbative result}
In the previous section we have obtained the particle production
probabilities up to second order in perturbation theory. 
This is in general a good approximation for small background
fileds, however even in those cases,  it does not contain all the 
information about
the particle production processes. In this
section we will study a particular configuration of vector and axial
fields for which it is possible to find an expression for the imaginary part 
which is non-perturbative in the electric field, in the limit 
$S^2<<E$ and $m^2S^2/E^2<1$.  

Let us start by introducing the operators $X_\mu$ and $P_\mu$ acting on
states $\vert x\rangle$ and $\vert p\rangle$ in the usual form:
$X_\mu\vert x\rangle=x_\mu\vert x\rangle$ and 
$P_\mu\vert p\rangle=p_\mu\vert p\rangle$. In addition $\langle x\vert P_\mu
\vert \phi\rangle=i\partial_\mu\langle x\vert \phi\rangle$. The conumutator
is given by  $[X_\mu,P_\nu]=
-ig_{\mu\nu}$ and $\langle p\vert x\rangle=e^{ipx}/(2\pi)^2$.

Following Itzykson and Zuber \cite{IZ} 
we recast the Dirac operator in (\ref{lag})
as:
\begin{eqnarray}
i\Dbar=(\Pbar-\Abar P_L-\Bbar P_R)
\end{eqnarray}
and taking the transpose we have:
\begin{eqnarray}
(\Pbar-\Abar P_L-\Bbar P_R)^t=-C(\Pbar-\Abar P_R-\Bbar P_L)C^{-1}
\end{eqnarray}
where $C=i\gamma^2\gamma^0$ is the charge conjugation matrix that
satisfies: $C\gamma_\mu C^{-1}=-\gamma_\mu^t$ and  
$C\gamma_5 C^{-1}=\gamma_5^t$. The effective action
in (\ref{ea}) can be written with
this notation as:
\begin{eqnarray}
W[A,B]=-i\Tr \log \left((\Pbar-\Abar P_L-\Bbar P_R-m+i\epsilon)
\frac{1}{\Pbar-m+i\epsilon}\right)
\end{eqnarray}
where we have explicitly introduced the normalization factor $N$
in the last term. The effective action can also be written in terms of the 
transposed operators:
\begin{eqnarray}
W[A,B]&=&-i\Tr \log \left((\Pbar-\Abar P_L-\Bbar P_R-m+i\epsilon)^t
\left(\frac{1}{\Pbar-m+i\epsilon}\right)^t\right)\nonumber \\
&=&-i\Tr \log \left((\Pbar-\Abar P_R-\Bbar P_L+m-i\epsilon)
\frac{1}{\Pbar+m-i\epsilon}\right)
\end{eqnarray}
adding both expression we get:
\begin{eqnarray}
2W[A,B]&=&-i\Tr \log \left((\Pbar-\Abar P_L-\Bbar P_R-m+i\epsilon)
(\Pbar-\Abar P_R-\Bbar P_L+m-i\epsilon)
\frac{1}{P^2-m^2+i\epsilon}\right)\nonumber \\
&=&-i\Tr \log\left[ \left((P_\mu-A_\mu P_R-B_\mu P_L)^2
-\frac{i}{4}(A_{\mu\nu}P_R+B_{\mu\nu}P_L)[\gamma^\mu,\gamma^\nu]
+m(\Abar-\Bbar)\gamma_5-m^2+i\epsilon\right)\right.\nonumber\\
&\cdot& \left.\frac{1}{P^2-m^2+i\epsilon}\right]
\end{eqnarray}
Finally we change to the vector and axial fields, the effective action
is then written as:
\begin{eqnarray}
2W[V,S]=-i\Tr \log\left[ \left((P_\mu-V_\mu-S_\mu\gamma_5)^2
-\frac{i}{4}(V_{\mu\nu}+S_{\mu\nu}\gamma_5)[\gamma^\mu,\gamma^\nu]
-m\Sbar\gamma_5-m^2+i\epsilon\right)\frac{1}{P^2-m^2+i\epsilon}\right]
\end{eqnarray}
Here $V_{\mu\nu}=\partial_\mu V_\nu-\partial_\nu V_\mu$ and $S_{\mu\nu}
=\partial_\mu S_\nu-\partial_\nu S_\mu$. Let us take the following background
fields: $V^\mu=(0,0,0,B x^1)$ and $S^\mu=(0,0,0,S)$ with $B$ and $S$ 
arbitrary constants. This choice correspons to a constant magnetic field $B$
along the $y$ axis and a constant axial field $S$ in the $z$ direction.
Obvisously the same result will be obtained if we choose the fields
in different spatial directions, provided they are orthogonal. By means
of the Schwinger proper-time integral we can write:
\begin{eqnarray}
2W[V,S]=-i\Tr\int_0^\infty \frac{ds}{s}e^{-is(m^2-i\epsilon)}
\left(\langle x\vert e^{is(P_0^2-P_1^2-P_2^2-(P^3-B X^1-S\gamma_5)^2
+\frac{i}{2}B[\gamma^1,\gamma^3]-mS\gamma_3\gamma_5)}\vert x\rangle
-\langle x\vert e^{isP^2}\vert x\rangle \right)
\end{eqnarray}
The action of the traslation operator will simplify this expression:
\begin{eqnarray}
(P^3-B X^1-S\gamma_5)^2 =
e^{-i\frac{P^1P^3}{B}}(-B X^1-S\gamma_5)^2 e^{i\frac{P^1P^3}{B}}
\end{eqnarray}
Let us now introduce complete sets of momentum eigenstates:
\begin{eqnarray}
2W[V,S]&=&-i\Tr\int_0^\infty \frac{ds}{s}e^{-is(m^2-i\epsilon)}\left[
\left(\frac{1}{(2\pi)^4}\int d^4p d\tilde p^1 e^{is(p_0^2-p_2^2)}e^{ip^3(
\tilde p^1-p^1)/B}
e^{-i(\tilde p^1-p^1)x^1}\right.\right.\nonumber \\
&\cdot&\left.\left.
\langle p^1\vert e^{is(-P_1^2-(-B X^1-S\gamma_5)^2
+\frac{i}{2}B[\gamma^1,\gamma^3]-mS\gamma_3\gamma_5)}\vert \tilde p^1
\rangle\right)
-\int \frac{d^4p d^4\tilde p}{(2\pi)^4}e^{i(\tilde p-p)x}
\langle p\vert e^{isP^2}\vert \tilde p\rangle \right]
\end{eqnarray}
Performing the integral in the $p^3$ variable and in the $p^0$ and
$p^2$ by means of:
\begin{eqnarray}
\int_{-\infty}^{\infty}dq e^{-isq^2}=\sqrt{\frac{\pi}{is}}
\end{eqnarray}
the above expression reduces to:
\begin{eqnarray}
2W[V,S]&=&i\Tr\int_0^\infty \frac{ds}{s}e^{-is(m^2-i\epsilon)}\left[
\frac{B}{8\pi^2}\int dp \langle p\vert e^{is(-P^2-(B X+S\gamma_5)^2
+\frac{i}{2}B[\gamma^1,\gamma^3]-mS\gamma_3\gamma_5)}\vert  p
\rangle-\frac{i}{(4\pi)^2s^2}\right]
\label{trazas}
\end{eqnarray}
where for simplicity we have denoted $X=X^1$ and $P=P_1$. 
The integral in $p$ together with the matrix trace can be considered as 
the trace of the evolution operator
corresponding to the  hamiltonian $H=H_0+H_1$ in ordinary
quantum mechanics with:
\begin{eqnarray}
H_0=\left(\begin{array}{cc}
-P^2-(BX+S)^2-B\sigma^2&0\\
0&-P^2-(BX-S)^2-B\sigma^2\
\end{array}\right)
\label{h0}
\end{eqnarray}
and 
\begin{eqnarray}
H_1=\left(\begin{array}{cc}
0&mS\sigma^3\\
mS\sigma^3&0\
\end{array}\right)
\end{eqnarray}
with $\sigma^i$ the corresponding Pauli matrices.
The problem of evaluating the effective action is thus reduced to the 
calculation of
the spectrum of the $H$ operator. However, since we cannot
 obtain such spectrum in an exact form, we will consider 
$H_1$ as a small perturbation.
With that purpose, we will assume that the contributions to the
spectrum coming from the perturbation are smaller than the
eigenvalues of $H_0$, i.e, $m^2S^2<B^2$.
We can then  apply the standard Kato theory for time-independent
perturbations in quantum mechanics \cite{galindo}. The $H_0$ operator is
made out of two shifted harmonic-oscillator hamiltonians with mass 
$M=1/2$, frequency $\omega=2B$ and  a new coupling to the magnetic
field. 
Its spectrum $\{\lambda_{n,i}^{(0)} \}$ and 
eigenfunctions $\{\psi_{n,i}^{(0)}\}$ with 
$\{ n=0..\infty, i=1..4\}$ can be easily obtained, they 
are given by:
\begin{eqnarray}\begin{array}{cc}
\lambda_{n,1}^{(0)}=E_n+B,\;\psi_{n,1}^{(0)}
=\frac{1}{\sqrt{2}}\left(\begin{array}{c}
i\phi_n\\ \phi_n\\0\\0\end{array}\right);&\lambda_{n,2}^{(0)}=E_n-B,
\;\psi_{n,2}^{(0)}=\frac{1}{\sqrt{2}}\left(\begin{array}{c}
\phi_n\\ i\phi_n\\0\\0\end{array}\right)\\
\lambda_{n,3}^{(0)}=E_n+B,\;\psi_{n,3}^{(0)}=\frac{1}{\sqrt{2}}
\left(\begin{array}{c}
0\\0\\i\hat\phi_n\\ \hat\phi_n\end{array}\right);&\lambda_{n,4}^{(0)}=
E_n-B,\;\psi_{n,4}^{(0)}=\frac{1}{\sqrt{2}}\left(\begin{array}{c}
0\\0\\\hat\phi_n\\ i\hat\phi_n\end{array}\right)
\end{array}
\end{eqnarray}
where $E_n=-2B(n+1/2)$ are the energy levels of the ordinary harmonic 
oscillator and:
\begin{eqnarray}
\phi_n(x)=\sqrt{\frac{\sqrt{B}}{\sqrt{\pi}2^nn!}}H_n\left(\sqrt{B}
\left(x+\frac{S}{B}\right)\right)e^{-\frac{B}{2}\left(x+\frac{S}{B}\right)^2}
;\;\;\hat\phi_n(x)=\sqrt{\frac{\sqrt{B}}{\sqrt{\pi}2^nn!}}H_n\left(\sqrt{B}
\left(x-\frac{S}{B}\right)\right)e^{-\frac{B}{2}\left(x-\frac{S}{B}\right)^2}
\end{eqnarray}
with $H_n$ the Hermite polynomials. Notice the different functional
form of these two functions, it reflects the fact that the two
harmonic oscillators in (\ref{h0}) are displaced in different ways.  
We will express the spectrum and
the eigenfuntions of the
complete hamiltonian $H$ as perturbative series: 
$\lambda_{n,i}=\sum_{p=0}\lambda_{n,i}^{(p)}$ and
$\psi_{n,i}=\sum_{p=0}\psi_{n,i}^{(p)}$. To first order in the perturbation,
the spectrum is given by the solutions of the equation \cite{galindo}:
\begin{eqnarray}
\det\vert\langle \psi_{n,i}\vert H_1\vert \psi_{n,j}\rangle
-(\lambda_{n,i}-\lambda_{n,i}^{(0)})\delta_{ij}\vert=0
\end{eqnarray}
The solution imply $\lambda_{n,i}^{(1)}=0, \; \forall n,i$, 
i.e. there is no first order 
correction to the energies. The second order contributions are given by:
\begin{eqnarray}
\lambda_{n,i}^{(2)}=\sum_{j\neq i}\frac{\vert\langle 
\psi_{n,i}^{(0)}\vert H_1\vert
\psi_{n,j}^{(0)}\rangle\vert^2}{\lambda_{n,i}^{(0)}-\lambda_{n,j}^{(0)}}
\end{eqnarray}
Although, the $H_0$ spectrum is double degenerated, it is possible to use
the above expression valid for non-degenerated spectra, since there is no 
contribution from states in the same multiplet. In any case, the explicit
calculation using Kato theory yields the same results.
The values of the second order perturbation can be evaluated in a 
straightforward
way, we get:
\begin{eqnarray}
\lambda_{n,1}^{(2)}=\lambda_{n,3}^{(2)}=
\frac{m^2S^2}{2B}\kappa_n^2,\;\;\lambda_{n,2}^{(2)}=
\lambda_{n,4}^{(2)}=-\frac{m^2S^2}{2B}\kappa_n^2
\end{eqnarray}
where 
\begin{eqnarray}
\kappa_n=\vert\langle \phi_n\vert\hat\phi_n\rangle\vert=e^{\frac{-S^2}{B}}
\vert L_n\left(2\frac{S^2}{B}\right)\vert
\end{eqnarray}
with $L_n$ the Laguerre polinomials. These polynomials are bounded
as $n$ grows, in fact asymptotically we have \cite{Grad}: 
$L_n(x)\sim\frac{1}{\sqrt{\pi}}e^{x/2}
x^{-1/4}n^{-1/4}\cos(2\sqrt{nx}-\pi/4)+\Od(n^{-3/4})$ for $x>0$ and therefore
$\kappa_n$ is bounded. In fact for $S^2/B<<1$, we can expand:
$\kappa_n^2=1-(S^2/B)(4n+2)+\Od(S^4/B^2)$. Therefore, as expected for small
values of the axial field with respect to the magnetic one, 
$\kappa_n^2=1$ is a good approximation.
Notice that, although the first correction grows like $n$, the growth 
must be controlled by higher order terms since the function
is bounded. Let us stress that we have two different parameters in
our problem: on one hand $S^2/B$ that we assume to be very small and
our perturbative parameter $m^2S^2/(2B^2)$.

Using (\ref{wave}) we can obtain the first correction to the eigenfunctions,
we get:
\begin{eqnarray}\displaystyle{\begin{array}{cc}
\vert \psi_{n,1}^{(1)}\rangle =\frac{imS}{2B}\kappa_n\vert 
\psi_{n,4}^{(0)}\rangle &
\vert \psi_{n,2}^{(1)}\rangle =\frac{imS}{2B}\kappa_n\vert 
\psi_{n,3}^{(0)}\rangle\\
\vert \psi_{n,3}^{(1)}\rangle =\frac{imS}{2B}\kappa_n\vert 
\psi_{n,2}^{(0)}\rangle &
\vert \psi_{n,4}^{(1)}\rangle =\frac{imS}{2B}\kappa_n\vert 
\psi_{n,1}^{(0)}\rangle\end{array}}
\end{eqnarray}
Evaluating now the third order corrections in perturbation
theory  from (\ref{easy}) they  again turn out to be zero.
This is also the case of the second order corrections to the eigenfuntions,
i.e. from (\ref{wave}) we get 
$\vert \psi_{n,i}^{(2)}\rangle=0\: \forall n,i$. 
Thus we can calculate fourth order corrections to the energies 
from (\ref{easy}) which are non-vanishing. Finally, 
we see that the fifth order correction  again vanishes. In conclusion,
our results for the perturbed spectrum, up to sixth order in perturbations
is given by:
\begin{eqnarray}
\lambda_{n,1}=\lambda_{n,3}&=&B\left(-2\left(n+\frac{1}{2}\right)
+1+\frac{m^2S^2}{2B^2}-\frac{m^4S^4}{8B^4}+
\Od\left(\frac{m^6S^6}{B^6}\right) \right)\nonumber \\
\lambda_{n,2}=\lambda_{n,4}&=&B\left(-2\left(n+\frac{1}{2}\right)
-1-\frac{m^2S^2}{2B^2}+\frac{m^4S^4}{8B^4}+
\Od\left(\frac{m^6S^6}{B^6}\right) \right)
\end{eqnarray}
Once we know the perturbed spectrum we can readily calculate the traces in 
(\ref{trazas}).  We will only perform the Dirac traces, but
not the functional trace that is equivalent to the integration
$\int d^4x$. Thus we obtain the result for the effective lagrangian $w$
that as expected does not depend on $x$:
\begin{eqnarray}
2w[V,S]&=&4i\int_0^\infty \frac{ds}{s^2}e^{-is(m^2-i\epsilon)}\left[
\frac{B}{8\pi^2}\cos\left(sB\left(1
+\frac{m^2S^2}{2B^2}-\frac{m^4S^4}{8B^4}+\Od\left(\frac{m^6S^6}{B^6}
\right)\right)\right)\sum_n e^{is(-2B(n+1/2))}\right.\nonumber \\
&-&\left.\frac{i}{(4\pi)^2s}\right]
\end{eqnarray}
Finally performing explicitly the addition of the series in $n$, we obtain:
\begin{eqnarray}
2w[V,S]=-\frac{4}{(4\pi)^2}\int_0^\infty \frac{ds}{s^2}
e^{-is(m^2-i\epsilon)}\left[
B\frac{\cos\left(sB\left(1
+\frac{m^2S^2}{2B^2}-\frac{m^4S^4}{8B^4}
+\Od\left(\frac{m^6S^6}{B^6}\right)\right)\right)}
{\sin(sB)}
-\frac{1}{s}\right]
\end{eqnarray} 
It can be seen that the result is purely real \cite{eaqed}, i.e, there is 
no particle production
in the presence of constant magnetic and axial fields. In addition, the
integral in $s$ is divergent in the ultraviolet limit $s\rightarrow 0$. 
As usual these divergencies has to be removed by adding suitable counterterms.
In order to obtain them, let us expand the part of the integrand in brackets
 around $s=0$, we have:
\begin{eqnarray}
B\frac{\cos\left(sB\left(1
+\frac{m^2S^2}{2B^2}-\frac{m^4S^4}{8B^4}+\Od\left(\frac{m^6S^6}{B^6}\right)
\right)\right)}
{\sin(sB)}
-\frac{1}{s}=B\left(\frac{1}{Bs}-\frac{sB}{2}-\frac{sm^2S^2}{2B}+
 \Od\left(\frac{m^6S^6}{B^6}\right)\right)-\frac{1}{s}+\Od(s^3)
\label{exp}
\end{eqnarray} 
The $\Od(s^3)$ give rise to finite contributions when integrated. The first
term is exactly cancelled by the  $1/s$ substraction. But we will have to 
include new counterterms proportional to $s(B^2+m^2S^2)$. Notice that 
these are
exactly the same operators that we found in the perturbative
calculation in section 3. In fact there is no contribution from 
$S^4$ operators due 
to an exact cancellation of  the different cuartic terms in the expansion in
(\ref{exp}). Since the theory is renormalizable,
it is neccessary that the same kind of cancellation operates
on the higher order terms in the r.h.s. of the above expression. 

 The presence of the axial field is known that does
not introduce any gauge anomaly in the electromagnetic
current \cite{DoMa0}, therefore the effective action
will be gauge invariant. As a consequence it should be built out of scalar
and gauge invariant functions. In our case, with constant magnetic
and torsion fields, the only possibilities are: $F_{\mu\nu}F^{\mu\nu}=
2(\vec B^2-\vec E^2)$ and $F_{\mu\nu}^*F^{\mu\nu}=
(4\vec B\cdot\vec E)^2$. Since the second term vanishes
in our case with constant magnetic or electric fields, the effective action
is invariant under the transformation $B\rightarrow -iE$ \cite{eaqed}. In
this way we can convert our results for constant magnetic field
to pure constant electric fields. Taking into account that now
$\sin(sB)\rightarrow -i\sinh(sE)$ and 
$\cos(sB(1+m^2S^2/(2B^2)-m^4S^4/(8B^4)))\rightarrow
\cosh(sE(1-m^2S^2/(2E^2)-m^4S^4/(8E^4)))$ and integrating using the 
residues technique we can
obtain the expression for the imaginary part of the effective lagrangian:
\begin{eqnarray}
p=2\im \; w[E,S]=\frac{1}{4\pi^3}E^2\sum_{n=1}^\infty\frac{(-1)^n}{n^2}
\cos\left(n\pi\left(1-\frac{m^2S^2}{2E^2}-\frac{m^4S^4}{8E^4}+
\Od\left(\frac{m^6S^6}{E^6}\right)\right)\right)e^{-\frac{m^2n\pi}{E}}
\end{eqnarray}
When $S=0$ we recover the original Schwinger result in (\ref{eres}).
Notice that in the absence of electric field there is no particle 
production even with non-vanishing axial field. When the mass is zero, 
we recover the usual result, i.e, axial fields only contribute in 
the massive case. In Figure. 1 the upper
curve  represent
the probability density $p$ as a function of $S$ for $E=0.1m^2$, this value for
the electric field ensures that the condition $S^2<E$ is satisfied 
for all the values of $S$ in the plot. 

\begin{figure}
\begin{center}
\mbox{\epsfysize=10cm\epsfxsize=10cm
\epsffile{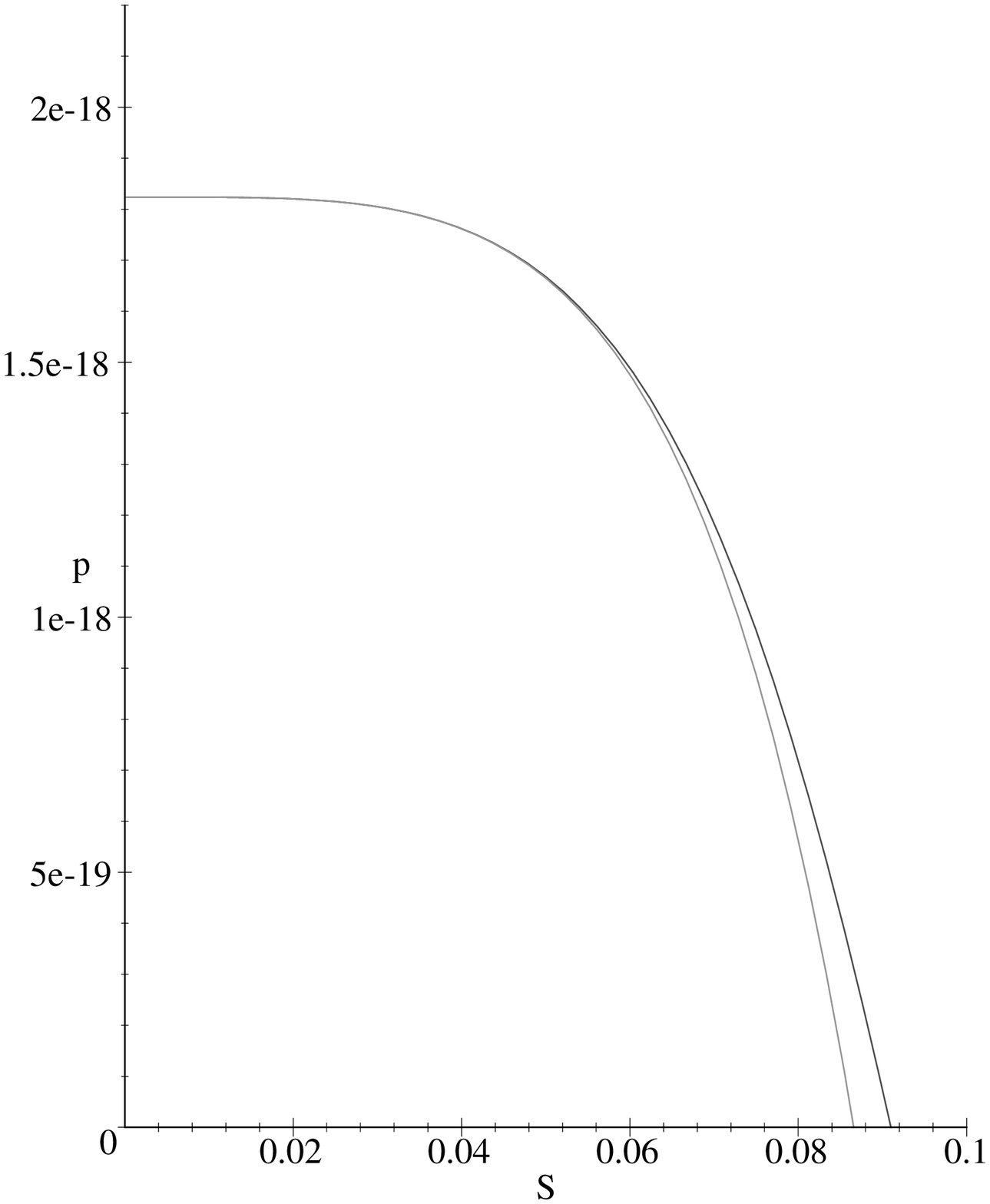}}

\end{center}
\vspace {-.8cm}
\leftskip 1cm
\rightskip 1cm
{\footnotesize
{\bf Figure 1.-}Probability  densities $p$ in units $m^4$ versus axial field
$S$ in units $m$, for an electric field $E=0.1m^2$. The upper curve
represents the perturbative calculation up to sixth order. The lower curve
is the resummation of the perturbative series estimated in the text.}

\end{figure}

We can try to extend further the above result by means of the following
observation. Since the only possible divergences in our model are
those mentioned before, it is neccessary that, when expanding the $\cos$
function in (\ref{exp}), the higher order terms in $S$ cancel, 
this implies:
\begin{eqnarray}
sB^2\left(1+\frac{m^2S^2}{2B^2}-\frac{m^4S^4}{8B^4}+\dots\right)^2=
s(B^2+m^2S^2)
\end{eqnarray}
Therefore we can obtain the complete result for the effective action to all
orders in perturbation theory in the limit $S^2<<B$. Performing the
rotation to electric fields, the result is given by :
\begin{eqnarray}
p=2\im\; w[E,S]=\frac{1}{4\pi^3}E^2\sum_{n=1}^\infty\frac{(-1)^n}{n^2}
\cos\left(n\pi\sqrt{1-\frac{m^2S^2}{E^2}}\right)e^{-\frac{m^2n\pi}{E}}
\end{eqnarray}
In Figure.1, this probability is represented by the lower curve.

We see that the effect of the axial field is to suppress particle production.
Eventually it could make it to vanish. The point of vanishing
$p$ should indicate the breakdown of the perturbative 
approximation since the probabilities should be positive. 
In order to check whether 
the perturbative calculation is valid up to the point in which $P$ vansihes,
we will study the convergence of the perturbative series. In particular, there
is a general result due to Kato \cite{Kato} that states that if there are two 
non-negative constants $a$ and $b$ such that: $\vert\vert H_1\vert\vert
\leq a\vert\vert H_0\vert \psi\rangle\vert\vert +b\vert\vert\vert 
\psi\rangle \vert\vert$ for all $\vert \psi \rangle\in D(H_0)$  
and if the operator $H_1$
is bounded then the eigenvalues perturbative series is absolutely
convergent if $2\vert\vert H_1\vert\vert<d$. With $d$ being the
distance from $\lambda_{n,i}^0$ to the rest of the spectrum of $H_0$.
In our case, we take $a=0$ and provided $\{\vert \psi_{n,i}\rangle\}$
is a complete set of eigenfunction we can
expand $H_1\vert \psi\rangle=
H_1\sum_{n,i}C_{n,i}\vert \psi_{n,i}\rangle$. It is then easy to show
that:
$\vert\vert H_1\vert\psi \rangle \vert\vert^2=m^2S^2\sum_{n,i}C_{n,i}^2
=m^2S^2$. Therefore if we take $b=mS$, since $H_1$ is bounded 
(it is constant)
and in our case $d=2B$, we have that the series is absolutely
convergent for $m^2S^2/(2B^2)<1/2$. Transforming
to the electrostatic case, if we look at the plot
we can realize that for those particular values, 
the point in which the curve
crosses the axis is in fact approximately signalling the breakdown of 
the perturbative series.   

\section{Conclusions}
In this work we have studied the production of massive fermions from  
a classical vector and axial-vector background. Using a perturbative
evaluation of the effective action we have obtained the contributions
to the imaginary part of the effective action
up to second order in the external fields. We have shown that in the reference 
frame for which $\vec k=0$, the production from purely spatial axial
fields is suppressed with respect to the that of the vector background. 
In addition, for purely temporal axial fields it is possible to create
particles
whereas this is not the case for vector fields. 

In the particular case of a small constant axial field and a constant
electric field, it is shown that a non-perturbative calculation can
be carried out when those fields are orthogonal. In this case, it is shown
that in the massless limit the axial field does not affect the production
from the electric field. However, in the massive case, the presence of the
axial background inhibits such production.

Finally, let us compare these result with the anisotropy 
damping phenomenon at the Planck era. 
As is well-known, the presence of small anisotropies
in the early universe can be damped in a few Planck times due to
the backreaction of the particles produced on the geometry.
An interesting possibility is that a similar mechanism could take 
place in the presence of some primordial torsion field. 
In this case, since
torsion can be generated by the initrinsic spin, this could happen when
the fermions are produced in some configuration such that the total spin
angular momentum 
of the system did not vanish.
The use of effective action methods for fermions could be extended to 
the production of higher spin fields, such as  gravitinos in a 
straightforward way. 
In addition,  it is also interesting to study not only the particle
production rate derived from the effective action, but also
the spectra and angular distribution of fermions produced. This could be
approached by means of the traditional Bogolyubov technique. Work
is in progress in this direction \cite{Anu}.

\section{Appendix}
In this Appendix we summarize the main formulae of the standard perturbation 
theory used in the text. 
Let us assume that the hamiltonian of theory can be decomposed in:
\begin{eqnarray}
H=H_0+H_1
\end{eqnarray}
As shown in the text, we denote by $\lambda_{n,i}^{(0)}$ and 
$\vert \psi_{n,i}^{(0)}\rangle$
the eigenvalues and eigenfunctions of $H_0$, which are assummed to
be known. The unperturbed spectrum is assumed to be
non-degenerated and discrete.
 The eigenfunctions and eigenvalues of the complete
hamiltonian $H$ are expanded in series: 
$\lambda_{n,i}=\sum_{p=0}\lambda_{n,i}^{(p)}$
and  $\vert \psi_{n,i}\rangle=\sum_{p=0}\vert\psi_{n,i}^{(p)}\rangle$
where:
\begin{eqnarray}
\lambda_{n,i}^{(p)}=\langle \psi_{n,i}^{(0)}\vert H_1\vert \psi_{n,i}^{(p-1)}
\rangle
\label{spec}
\end{eqnarray}
and
\begin{eqnarray}
\vert\psi_{n,i}^{(p)}\rangle=S_{n,i}\left(H_1\vert\psi_{n,i}^{(p-1)}\rangle
-\sum_{k=1}^{p-1}\lambda_{n,i}^{(k)}\vert\psi_{n,i}^{(p-k)}\rangle\right)
\label{wave}
\end{eqnarray}
with 
\begin{eqnarray}
S_{n,i}=\sum_{j\neq i}\frac{\vert \psi_{n,j}^{(0)}\rangle
\langle\psi_{n,j}^{(0)}\vert}{\lambda_{n,i}^{(0)}-\lambda_{n,j}^{(0)}}
\end{eqnarray}
Some simplified formulae for the lowest order terms in the 
spectrum are given by:
\begin{eqnarray}
\lambda_{n,i}^{(1)}&=&\langle \psi_{n,i}^{(0)}\vert H_1\vert \psi_{n,i}^{(0)}
\rangle \nonumber \\
\lambda_{n,i}^{(2)}&=&\sum_{j\neq i}\frac{
\vert\langle \psi_{n,i}^{(0)}\vert H_1\vert \psi_{n,j}^{(0)}
\rangle\vert ^2}{\lambda_{n,i}^{(0)}-\lambda_{n,j}^{(0)}}\nonumber \\
\lambda_{n,i}^{(3)}&=&\langle \psi_{n,i}^{(1)}\vert H_1-\lambda_{n,i}^{(1)}
\vert \psi_{n,i}^{(1)}
\rangle 
\nonumber \\
\lambda_{n,i}^{(4)}&=&\langle \psi_{n,i}^{(1)}\vert H_1-\lambda_{n,i}^{(1)}
\vert \psi_{n,i}^{(2)}\rangle-\lambda_{n,i}^{(2)}\langle \psi_{n,i}^{(1)}\vert
 \psi_{n,i}^{(1)}\rangle
\nonumber\\
\lambda_{n,i}^{(5)}&=&\langle \psi_{n,i}^{(2)}\vert H_1-\lambda_{n,i}^{(1)}
\vert \psi_{n,i}^{(2)}\rangle-2\lambda_{n,i}^{(2)}\re
\langle \psi_{n,i}^{(1)}\vert
 \psi_{n,i}^{(2)}\rangle-\lambda_{n,i}^{(3)}\langle \psi_{n,i}^{(1)}\vert
 \psi_{n,i}^{(1)}\rangle
\label{easy}
\end{eqnarray}

{\bf Acknowledgements:} I thank Ed. Copeland and A. Dobado for
useful suggestions. This work has been partially supported by
the Ministerio de Educaci\'on y Ciencia (Spain) (CICYT AEN96-1634).
The author also acknowledges support from SEUID-Royal Society.

\end{document}